# Hyper-Connected Transformer Network for Multi-Modality PET-CT Segmentation

Lei Bi, Michael Fulham, Shaoli Song, David Dagan Feng, *Fellow, IEEE*, Jinman Kim, *Member, IEEE*

*Abstract*—[18F]-Fluorodeoxyglucose (FDG) positron emission tomography – computed tomography (PET-CT) has become the imaging modality of choice for diagnosing many cancers. Co-learning complementary PET-CT imaging features is a fundamental requirement for automatic tumor segmentation and for developing computer aided cancer diagnosis systems. In this study, we propose a hyper-connected transformer (HCT) network that integrates a transformer network (TN) with a hyper connected fusion for multi-modality PET-CT images. The TN was leveraged for its ability to provide global dependencies in image feature learning, which was achieved by using image patch embeddings with a self-attention mechanism to capture image-wide contextual information. We extended the single-modality definition of TN with multiple TN based branches to separately extract image features. We also introduced a hyper connected fusion to fuse the contextual and complementary image features across multiple transformers in an iterative manner. Our results with two clinical datasets show that HCT achieved better performance in segmentation accuracy when compared to the existing methods.

*Clinical Relevance*—We anticipate that our approach can be an effective and supportive tool to aid physicians in tumor quantification and in identifying image biomarkers for cancer treatment.

## I. INTRODUCTION

Multi-modality positron emission tomography and computed tomography (PET-CT) using 18F-Fluorodeoxyglucose (FDG) has become the preferred method for evaluating cancers such as lung cancers and soft-tissue sarcomas [1]. This is because the combination of PET and CT in one scan provides both the increased sensitivity (from PET) in identifying abnormal activity and the detailed anatomy information (from CT). Computer aided diagnosis (CAD) is a growing field of research aimed at using computer-generated results to provide additional support for physician's image interpretation, enhance diagnostic results, and shorten image interpretation time [2]. Automated PET-CT tumor segmentation is regarded as the first step in implementing CAD systems and the underlying aim is to co-learn available complementary features from the PET and CT images, and then use the derived image features to separate the tumors from the surrounding background. Motivated by this need, various segmentation methods have been developed for PET-CT.

Data driven deep learning methods are driving significant advancements and novel discoveries in all areas of medical image analysis, including disease classification, segmentation, and retrieval. In segmentation, the most advanced techniques are based on fully convolutional networks (FCNs) [3]. FCN uses a combination of a convolutional neural network (CNN) for feature extraction and a CNN decoder for upsampling to produce the final segmentation results. These networks have been applied to multi-modality PET-CT image segmentation in various ways, including multi-view approach [4], using multiple FCNs for each modality [5] and using PET images as an attention map for segmentation from CT images [6]. However, these FCN-based methods can struggle with capturing long-range dependencies in the image, as their convolutional kernels have limited receptive fields, which can cause the network to overlook global context and focus solely on local patterns. Efforts have been made to improve the long-range dependency through stacking convolutional layers and using attention modules, but the conventional FCN architecture still restricts the receptive fields [7].

Recently, transformer networks (TNs), deep learning models designed for sequence data, have emerged as the leading technology for various image analysis tasks, including image classification and object detection [7]. This success is due to the integration of image patch embeddings with a self-attention mechanism, allowing for the creation of long-range dependencies and capturing of global context. As a result, TN-based methods for medical image segmentation are also gaining popularity. Chen et al. [8] introduced the TransUNet architecture, which combines a classic U-shaped UNet architecture with a TN-based encoder. Zhang et al. [9] proposed to integrate features from both FCN and TN for polyp segmentation in endoscopy images. Valanarasu et al. [10] employed two TNs to segment both ultrasound and histopathology images, with one TN designed to capture global-level features and the other to extract local-level (image patch) features. However, all these TN based segmentation methods were designed for single modality image segmentation. A straightforward translation of single-modality TN to multi-modality TN is to conduct early-fusion, where multi-modality image features are fused priori to FCN, or late-fusion, where the learned resultant features are fused. However, early- and late-fusion methods have limited freedom to fuse multi-modality image features and tend to dismiss complementary correlations across different modalities.

### A. Our Contribution

We propose a hyper-connected transformer (HCT) for automatic PET-CT tumor segmentation. HCT extends the

This work was supported in part by Australian Research Council (ARC) grant (DP200103748).

Lei Bi is with the Institute of Translational Medicine, National Center for Translational Medicine, Shanghai Jiao Tong University, Shanghai, China. (corresponding e-mail: lei.bi@sjtu.edu.cn).

David Dagan Feng and Jinman Kim are with the School of Computer Science, University of Sydney, NSW, Australia.

Michael Fulham is with the Department of Molecular Imaging, Royal Prince Alfred Hospital, NSW, Australia.

Shaoli Song is with the Department of Nuclear Medicine, Fudan University Shanghai Cancer Center, Fudan University, Shanghai, China

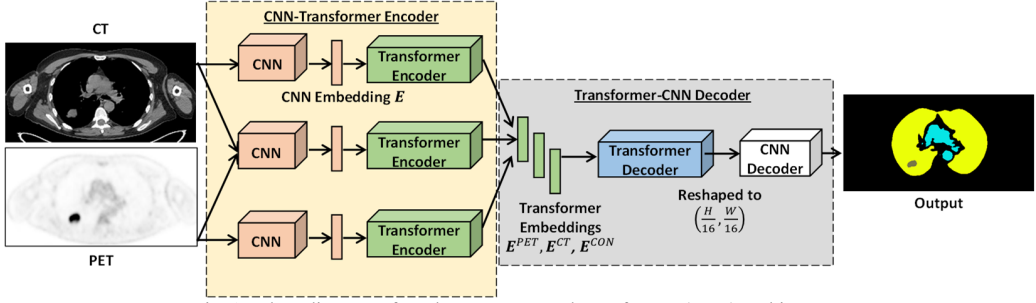

Fig. 1. Flow diagram of our hyper-connected transformer (HCT) architecture.

definition of TN with multiple TN based branches to separately extract image features from PET-CT images. These complementary image features are then fused in a progressive manner across multiple transformers for segmentations. Our HCT adds the following contributions to the current knowledge by our proposition: (1) Our HCT provides the flexibility to learn complementary features in a long-range dependency and overcome the limitations that current FCN based methods have in segmenting PET-CT; (2) We propose to separately extract features and fuse the extracted PET and CT image features with transformers, which allows continuous fusion of the complementary PET and CT image characteristics, and minimizes the loss of information during early- or late-fusion; and (3) We propose a TN based decoder to embed the separately extracted PET and CT features; the decoder parses and prioritizes the segmentation relevant features across the PET and CT, and ensures that challenging tumors e.g., tumors with heterogeneous textures can be detected.

## II. Materials and Methods

### A. Overview of the Proposed Method

Our hyper-connected transformer (HCT), as shown in Fig. 1, consists of a CNN-TN encoder (CNN-TN-E) and a TN-CNN decoder (TN-CNN-D). CNN-TN-E has three branches to separately process PET, CT and concatenated PET-CT images. Each branch also employs a hybrid approach to encode image features with CNNs into embeddings $E$. The encoded image embeddings are then processed by the transformer encoders to learn complementary features in a long-range dependency between the PET, CT and concatenated PET-CT images (respectively of $E^{PET}$, $E^{CT}$, $E^{CON}$). The learned embeddings are fused by the transformer decoder to identify segmentation relevant features, which are then reshaped to a 2D feature map. Finally, a CNN is used to upsample the fused features and to output the segmentation results.

### B. CNN-Transformer Network Encoder (CNN-TN-E)

CNN-TN-E consists of a PET image branch, a CT image branch and a hyper-branch. The PET and CT image branches were used to derive feature encodings for PET and CT images. The hyper-branch was used to process the concatenated PET and CT images.

To prepare the input to the transformer encoder, we firstly projected the input image $I \in \mathbb{R}^{H \times W \times C}$ into a 1-dimensional (1D) feature vector, where $H$ represents the height of the input image, and $W$ represents the width and $C$ is the number of channels. $C$ was set to 1 for the input PET or CT images and was set to 2 for the concatenated input. To minimize the required computational complexity of the transformer encoder, we used CNN to produce 2D image feature maps at a size of $(\frac{H}{16}, \frac{W}{16}, C)$. We used the residual neural network (ResNet) [11] (ResNet-50) as the CNN backbone.

To encode the spatial information, a positional embedding was learned which was then added to the extracted image feature maps. We firstly flattened and projected the feature map into a $D$-dimensional embedding space $E_0 = (\frac{H \times W}{16 \times 16}, D) = \{e_1, e_2, ..., e_D\}$. After that, we calculated a positional embedding $P = \{p_1, p_2, ..., p_D\}$ for every embeddings based on a truncated normal distribution. The final embedding was defined as $E = \{e_1 + p_1, e_2 + p_2, ..., e_D + p_D\}$.

Our transformer encoder has multiple transformer modules, representing transformer encoder with different depths $T$ [12]. Each transformer module has multi-head self-attention (MSA), multi-layer perceptron (MLP) and layer normalization (LN) blocks to extract image features from the embeddings.

### C. Transformer Network-CNN Decoder (TN-CNN-D)

At the end of our CNN-TN-E, we generated three separate embeddings $E^{PET}$, $E^{CT}$, $E^{CON}$ representing image features extracted from PET, CT and concatenated images. Similar to the transformer encoders, the transformer decoder also has multiple transformer modules representing transformer decoder with different depths. The transformer module was purposely designed to take concatenated embeddings and then to use the positional embedding to learn spatial information within the embeddings and across PET and CT images, which can be defined as:

$$E = \{e_1^{PET} + p_1, ..., e_D^{PET} + p_D, e_{D+1}^{CT} + p_D, ..., e_{2D}^{CT} + p_{2D}, e_{2D+1}^{CON} + p_{2D+1}, ..., e_{3D}^{CON} + p_{3D}\} \quad (1)$$

The output embedding from the transformer decoder was reshaped into a feature map at a size of $(\frac{H}{16}, \frac{W}{16})$ and were further processed by two convolutional (Conv) layers. We followed the design of the traditional FCN architecture and added the CNN produced feature map. Specifically, the feature map derived from the CNN encoder was processed by a Conv layer and fused to the linearly interpolated transformer produced feature map. The final segmentation was obtained by using a Conv layer and a pixel-wise softmax operation.

### D. Implementation Details

PET images were normalized to standardized uptake values (SUVs) and were then normalized to [0, 15]. CT images were converted to Hounsfield units and were then normalized to [-160, 240]. Our HCT was developed with PyTorch and the backbone ResNet was fine-tuned from a pre-trained ImageNet model. At the training stage, we applied online data augmentation (random cropping, resizing and flipping) and used the pixel-wise cross-entropy loss to train HCT for 100 epochs on a NVIDIA 24GB 3090 GPU. The batch size was set to 2. The learning rate was set to $1e^{-4}$ and was decayed with a poly strategy. The default of HCT depth was set to 4 and the embedding was set to 256. The Adam approach was used as the optimizer with an epsilon value of $1e^{-8}$ and weight decay was set to $1e^{-4}$.

### E. Materials

We used one non-small cell lung cancer (NSCLC) and one soft-tissue sarcoma (STS) datasets [13] for evaluation. The NSCLC scans were acquired from the Department of Nuclear Medicine at Fudan University Shanghai Cancer Center, Shanghai, China (denoted as FD dataset). There were 70 patients and all scans were acquired on a Biograph 16-slice TruePoint scanner (Siemens Healthineers, Hoffman Estates, Chicago, IL, USA). Both PET and CT volumes were reconstructed with the same number of slices. Tumors were delineated independently by two radiologists using ITK-SNAP software (Version 3.6, United States). In case of inconsistency among two radiologists, especially for the contours adjacent to the mediastinum, chest wall, and blood vessels, contours were confirmed from a senior radiologist.

The STS dataset was acquired from McGill University Health Centre, Quebec, Canada, comprising 51 PET-CT scans acquired from the McGill University Health Centre. All scans were acquired on a Discovery ST scanner (GE Healthcare, Waukesha, WI). The tumor contours were manually delineated by an expert radiation oncologist on the T2-weighted fat-suppression MRI scans, then propagated to PET-CT images via rigid registration.

In all two datasets, as a pre-processing step, PET images were linearly interpolated to the same size as the CT images. Upsampling of PET images was chosen over downsampling of CT images to avoid losing pixel information.

### F. Experimental Setup

A 5-fold cross-validation evaluation protocol was conducted for each dataset. For example, with our FD dataset having 70 patient studies, image slices from 56 studies were used as the training set and image slices from the remaining 14 studies were used as the test set. The experiments we conducted were as follows: We evaluated our HCT with different multi-modality fusion strategies with the use of FCN and TN. We used ResNet-50 as the network backbone: (1) EF (early fusion) – multi-modality inputs fused at an early stage of the network; (2) LF (late fusion) – multi-modality inputs were separately processed by multiple networks with the output results been fused; and (3) HF (hyper fusion) – multi-modality image features fused within the network architecture.

We compared our HCT to current PET-CT segmentation methods and the included methods were: (1) WNet [5] – two VNets (named as WNet) for iterative segmentation, where the first VNet was applied to the input CT images and the segmentation results were refined with the second VNet together with PET and CT images as the input; (2) a co-learning [14] – a two branched UNet was implemented to extract and fuse PET and CT features across multiple UNet blocks; (3) UNet [15] with an early-fusion strategy; (4) RFN [16] – recurrent fusion network, where cascaded FCNs were used for fusing the PET and CT imaging features; (5) MSAM [6] – multi-modal spatial attention module, which uses the PET images as an attention map to guide the tumor segmentation on CT images; (6) TN [12] – a conventional transformer based segmentation with an early-fusion strategy; (7) PVT [17] – pyramid vision transformer where the concatenated PET and CT images were used as the input.

The included evaluation metrics are: Dice score (DSC), precision (Pre.), sensitivity (Sen.), specificity (Spe.) and precision-recall (PR) curve.

## III. RESULTS

### A. Segmentation with Various Fusion Strategies

Fig. 2 shows the segmentation results with different fusion strategies measured via the precision-recall (PR) curve. The results indicate that TN based methods can achieve higher segmentation accuracy when compared with the FCN based counterpart across different fusion strategies. We attribute this to the use of the transformer for leveraging image patch embeddings with self-attention mechanism to establish a long-range dependency which is able to capture global context, and thus aid in the removal of the false positive regions.

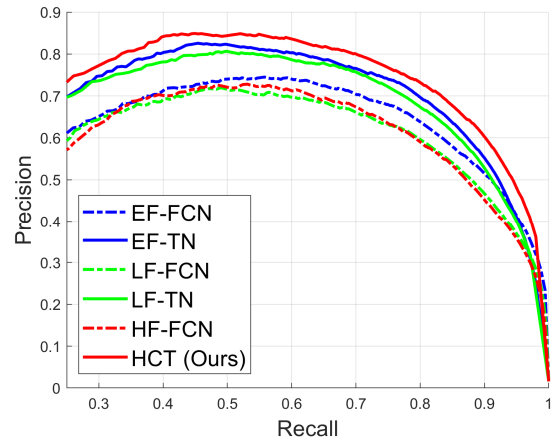

Fig. 2 Precision-recall (PR) curves on the FD dataset with different fusion methods.

Compared to the EF-TN, the LF-TN separately processed the multi-modality inputs with the output results being fused just prior to the output. This resulted in the complementary image features not being learnt in a joint manner. With the second-best performing EF-TN, segmentation results tended to focus on the PET image features while dismissing the complementary CT features; in this process, the complementary CT features could already be lost during early fusion. In contrast, our HCT separately processed and then co-learned PET and CT image features, and minimized the loss of information during the early-fusion process. In addition, our transformer based decoder allows for the fusion of the separately extracted PET and CT image features and prioritizes the segmentation relevant image features. Therefore, compared to LF-TN method, which fused image

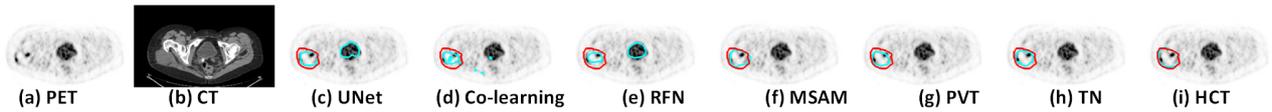

(a) PET   (b) CT   (c) UNet   (d) Co-learning   (e) RFN   (f) MSAM   (g) PVT   (h) TN   (i) HCT

Fig. 3. Segmentation results with one soft-tissue sarcoma example shown on transaxial PET (a) and CT (b) image slices. The segmentation results are shown in the following columns in (c) to (i). The red contour outlines the 'ground truth' segmentation; the cyan contour outlines the results from the comparison methods.

features just prior to the final output, our HCT exhibited flexibility in retaining important PET and CT image features for segmentation.

### B. Comparing to the State-of-the-art

Table I and Fig. 3 show the comparison results of our HCT to the state-of-the-art multi-modality segmentation methods. TN achieved 1.82% better in DSC on the FD dataset and 2.04% improvement in DSC on the STS dataset when compared to the second-best performing FCN based method of RFN. Our HCT further improved the TN, measured in DSC, by an average of 2.68% with FD and 1.4% with STS.

The improvement of MSAM over Co-learning is likely attributed to the use of multi-modality spatial attention module, which allows it to leverage the PET images as an attention map to guide the tumor segmentation on CT images. RFN further improved MSAM by using an iterative approach to progressively refine the segmentation results. Consistent with our ablation results, these FCN based methods are inherently restricted by the limited receptive field, which prohibited these methods from learning the global context.

The better performance of TN and PVT over RFN we suggest relates to the use of the transformer to capture the global context. TN shows competitive segmentation performance to our HCT method on the FD and STS datasets. However, both the TN and the PVT are designed for single modality image segmentation, where early fusion (concatenation) was applied for PET-CT image segmentation. Consequently, the useful complementary information could be lost during early fusion and resulted in limited segmentation performance in segmenting challenging tumors e.g., tumors with low-contrast to the background and tumors with heterogeneous textures (Fig. 3). In contrary, our HCT fused the PET, CT and concatenated PET-CT image features across multiple transformer modules; the fusion of various image features enabled the appearance of the results to be in an agreement with different imaging modalities.

TABLE I. COMPARISON WITH THE STATE-OF-THE-ART SEGMENTATION METHODS ON THE FD AND THE STS DATASETS.

| Datasets | FD | STS | | | |
|---|---|---|---|---|---|
| Metrics | DSC | DSC | Pre. | Sen. | Spec. |
| UNet [15] | 49.78 | 59.63 | 64.50 | 64.49 | 99.65 |
| WNet [5] | 50.54 | - | - | - | - |
| Co-learning [14] | 52.49 | 59.07 | 68.95 | 60.06 | 99.65 |
| RFN [16] | 67.75 | 62.92 | 69.27 | 66.07 | 99.69 |
| MSAM [6] | - | 62.26 | 69.00 | 64.74 | **99.74** |
| PVT [17] | 66.87 | 65.83 | **72.92** | 67.65 | 99.69 |
| TN [12] | 69.57 | 64.96 | 71.57 | 66.37 | **99.74** |
| HCT | **72.25** | **66.36** | 71.45 | **69.93** | 99.69 |

## IV. CONCLUSION

We introduced a hyper-connected fusion network to fuse the contextual and complementary image features from PET-CT through multiple transformer networks. This approach which minimized the risk of losing information during early- and late-fusion. Our results with two clinical datasets showed that HCT outperformed comparative state-of-the-art methods across different PET-CT datasets, which suggests that our method is more generalizable than the existing methods.